\documentclass[a4paper,12pt]{article}
\usepackage{graphicx,epsf}

\begin{document}
\title{Conserved non-linear  quantities in cosmology}
\author{David Langlois$^{1,2}$, Filippo Vernizzi$^3$\\
{\small {}}\\
{\small ${}^1${\it APC (Astroparticules et Cosmologie),}}\\
{\small {\it
UMR 7164 (CNRS, Universit\'e Paris 7, CEA, Observatoire de
Paris)}}\\
{\small {\it  11 Place Marcelin Berthelot, F-75005 Paris, France;}}\\
{\small ${}^2${\it GReCO, Institut d'Astrophysique de Paris, CNRS,}}\\
{\small {\it 98bis Boulevard Arago, 75014 Paris, France;  }}\\
{\small {\it and}}\\
{\small ${}^3${\it Helsinki Institute of Physics, P.O. Box 64,}}\\
{\small {\it FIN-00014 University of Helsinki - Finland}}\\
}

\date{\today}
\maketitle

\def\beq{\begin{equation}}
\def\eeq{\end{equation}}
\newcommand{\bea}{\begin{eqnarray}}
\newcommand{\eea}{\end{eqnarray}}
\def\bi{\begin{itemize}}
\def\ei{\end{itemize}}
\def\Tdot#1{{{#1}^{\hbox{.}}}}
\def\Tddot#1{{{#1}^{\hbox{..}}}}
\def\D{{D}}
\def\d{{\delta}}
\def\T{{\bf T}}
\def\perp{n}
\def\R{{\cal K}}
\def\L{{\cal L}_u}
\def\HH{{\cal H}}

\begin{abstract}
We  give a detailed and improved presentation of our recently
proposed formalism for non-linear perturbations in cosmology,
based on a covariant and fully non-perturbative approach. We work,
in particular, with a covector combining  the gradients of  the
energy density and of the local number of e-folds to obtain a
non-linear generalization of the familiar linear uniform density
perturbation. We show that this covector obeys a remarkably simple
conservation equation which is exact, fully non-linear and valid
at all scales. We relate explicitly  our approach to the
coordinate-based formalisms for linear perturbations and for
second-order perturbations. We also consider other quantities, 
 which 
are conserved on sufficiently large scales for adiabatic
perturbations, and discuss the issue of gauge invariance.
\end{abstract}

\newpage

\section{Introduction}
The relativistic theory of cosmological perturbations is an
essential tool to analyze cosmological data such as the Cosmic
Microwave Background (CMB) anisotropies, and thus to connect
 the scenarios of the early universe, such as inflation,
to  cosmological observations. Because the temperature
anisotropies of the CMB are so small ($\delta T/T \sim 10^{-5}$),
considering only {\it linear} perturbations is an excellent
approximation. This is  why most of the efforts devoted to the
theory of cosmological perturbations have dealt with linear
perturbations
\cite{LK,Bardeen:1980kt,Kodama:1985bj,Mukhanov:1990me,Durrer}.

There are however some issues where   one must take into account
the  non-linear aspects of cosmological perturbations. A good
example is the  study of inhomogeneities on scales much larger
than the Hubble radius, where the Universe could strongly deviate
from a Friedmann-Lema\^{\i}tre-Robertson-Walker (FLRW) geometry.
Another motivation, which has seen renewed interest recently, is
to investigate the predictions of early universe models for
primordial non-Gaussianity, with the hope to be able to detect
this non-Gaussianity in the CMB data. This investigation requires
the study  of relativistic cosmological perturbations {\it beyond
linear order} \cite{Bartolo:2004if,Bartolo:2004ty,Bartolo:2005fp}.

For small perturbations, as a  first step beyond linear order, one can
consider second-order relativistic perturbations. This is enough
if one wants to compute the bispectrum of perturbations, an
indicator of the presence of non-Gaussianities.
In the context of general relativity, dealing with
second-order perturbations is already an impressive task.
The usual treatment introduces coordinates and the
metric perturbations are then formally assumed to be the sum of a
first-order quantity and of a second-order quantity
\cite{Bruni:1996im,Bruni:1996rg,Acquaviva:2002ud,Malik:2003mv,Noh:2004bc,Vernizzi:2004nc,Malik:2005cy}.
One can treat similarly the fluid quantities and then write down
Einstein's equations up to second order.   As one can
imagine, this approach is rather cumbersome.

Other recent approaches of non-linear perturbations
\cite{Rigopoulos:2003ak,Kolb:2004jg,Lyth:2004gb} (see also
\cite{Afshordi:2000nr}) are based on the long wavelength
approximation \cite{Salopek:1990jq,Comer:1994np,Deruelle:1994iz},
which at lowest order is related to the so called separate
universe picture that represents our universe, on scales larger
than the Hubble radius, as juxtaposed FLRW universes with
slightly different scale factors \cite{Starobinski,Sasaki:1995aw}.

In the present work, we present the details of a different
approach, which we proposed recently \cite{lv05a}, based on a
purely geometrical description of the perturbations.  Our
approach is inspired by the so-called covariant formalism for
cosmological perturbations introduced by Ellis and Bruni
\cite{Ellis:1989jt}. Although the linearized version of this
formalism is often used and it has also been used at second order
(see e.g. \cite{clarkson}), our approach is fully non-perturbative
and thus not limited to second-order perturbations. The equations
and the variables used in our approach encode the full
non-linearity of cosmological perturbations on all scales.
Although, at linear order, the covariant formalism is {\it
computationally} equivalent to the much more used coordinate
approach, the covariant approach turns out to be more efficient when
one goes beyond linear order.

In particular, we show that it is possible to define, in a
geometric way, the generalizations of quantities widely used in
the linear theory because they are {\it conserved} on large
scales. These quantities are covectors which obey conservation
equations with respect to the directional derivation (Lie
derivation) along the comoving worldlines. We give the full
non-linear equations that govern these quantities. To emphasize
the efficiency of our approach with respect to the more
traditional coordinate-based approach, we show how gauge-invariant
quantities, recently derived in the literature, can be obtained
from our geometric quantities in a straightforward derivation. The
particular case of scalar fields perturbations will be considered
in a separate publication.

This paper is organized as follows. In the next section, we give a
brief overview of the covariant formalism for cosmological
perturbations due to Ellis and Bruni. In Sec.~3, we show how the
energy conservation equation leads naturally to a covector which
obeys a conservation equation. In Sec.~4, we show explicitly the
connection between our approach and the familiar,
coordinate-based, linear theory. In Sec.~5, we study second order
perturbations. In Sec.~6, we consider other non-linear conserved
quantities. In Sec.~7, we discuss the gauge
invariance of our variables and finally, in Sec.~8, we summarize
and conclude.

\section{Covariant formalism} \label{sec:choices}

In this section, we briefly review the basic ideas of the
covariant approach developed by Ellis, Bruni and collaborators
\cite{Ellis:1989jt,Ellis:1989ju,Bruni:1992dg}, and based on
earlier works by Hawking \cite{Hawking:1966qi} and Ellis
\cite{ellis}. For simplicity, we consider the universe filled with
a single perfect fluid, although most of what we will say applies
to any independent (i.e., non-interacting with other fluids)
perfect fluid, whether it is or not surrounded by other fluids.
The case of 
an imperfect fluid, with non-vanishing heat flow and anisotropic stress, 
will be considered in a separate publication.
The fluid we consider can be characterized by a comoving
four-velocity $u^a = d x^a/d \tau$ ($u_a u^a =-1$), where $\tau$
is the proper time along the flow lines, a proper energy density
$\rho$ and a pressure $P$.

The energy-momentum tensor associated to the perfect fluid is given by
\beq
T^a_{\ b}=\left(\rho+P\right) u^a u_b+Pg^a_{\ b}. \label{emt}
\eeq
To fully characterize the fluid, one needs an equation of state
relating $P$ to $\rho$ and, possibly, to other physical quantities
if the fluid is not barotropic.

The spatial projection tensor orthogonal to the fluid velocity
$u^a$ is defined by
\beq
h_{ab}=g_{ab}+u_a u_b, \quad \quad (h^{a}_{\ b} h^b_{\ c}=h^a_{\
c}, \quad h_a^{\ b}u_b=0).
\eeq
It is also useful to introduce the familiar decomposition
\beq
\nabla_b u_a=\sigma_{ab}+\omega_{ab}+{1\over 3}\Theta
h_{ab}-\dot{u}_a u_b, \label{decomposition}
\eeq
with the (symmetric) shear tensor $\sigma_{ab}$, and the
(antisymmetric) vorticity  tensor $\omega_{ab}$; the volume
expansion $\Theta$, is defined by
\beq
\Theta \equiv \nabla_a u^a,
\eeq
while $\dot u^a$ is the acceleration, with the dot denoting the
covariant derivative projected along $u^a$, i.e., $\dot{} \equiv  u^a \nabla_a
$.

The integration of $\Theta$ along the fluid world lines with
respect to the proper time $\tau$ is
\beq
\label{alpha_def} \alpha \equiv {1\over 3}\int d\tau \, \Theta,
\eeq
and can be used to define,  for each observer comoving with the
fluid, a local scale factor $S=e^\alpha$. It follows that
\beq
\Theta = 3 \dot\alpha \equiv 3 u^a\nabla_a\alpha  .
\eeq
Note that $\alpha$ is defined up to an integration constant for
{\it each} fluid world line. A convenient way to specify this
integration constant is to introduce some reference hypersurface
on which $\alpha=0$. The arbitrariness in the choice of the
integration constant along each world-line is thus translated in
the arbitrariness in the choice of this reference hypersurface.

We now wish to define, in a geometrical way, quantities that can
be interpreted as perturbations with respect to the FLRW
configuration. Since we work directly with the clumpy spacetime,
an obvious difficulty is  how to separate homogeneous quantities
from their perturbations. A way to get around this difficulty was
first suggested by Ellis and Bruni \cite{Ellis:1989jt}. They
introduced {\it spatial} projections (defined as projections
orthogonal to the four-velocity $u^a$) of the covariant derivative
of various scalar quantities: of the energy density,
\beq
X_a=h_a^{\ b}\nabla_b\rho\equiv D_a\rho, \label{X_def}
\eeq
of the pressure
\beq
Y_a\equiv D_a P,
\label{Y_def}
\eeq
and  the volume expansion,
\beq
Z_a\equiv D_a \Theta.
\label{Z_def}
\eeq
Here, following \cite{lv05a},
we also introduce the spatial gradient  of the integrated expansion,
\beq
W_a\equiv D_a \alpha. \label{W_def}
\eeq
Note that these definitions are purely geometrical and depend only
on the (physical) four-velocity $u^a$. All these quantities
automatically vanish in a strictly FLRW spacetime: in this sense
we call them {\it perturbations}. However, they are fully
non-perturbative quantities, not restricted to linear order in a
perturbation expansion. The last quantity, $W_a$, corresponds to
the spatial gradient of the local number of e-folds $\alpha$. It
was introduced in the non-perturbative approach in our previous
work \cite{lv05a}, because it plays a crucial r\^ole in our
approach, replacing the familiar spatial curvature perturbation,
similarly to the separate universe picture
\cite{Sasaki:1995aw,Wands:2000dp}. Note that a quantity similar to
$W_a$ has already been used in the linearized covariant theory in
\cite{Lewis}.

\section{Generalized conserved quantities} \label{sec:evolution}
Starting from the above  definitions, we will now introduce a
quantity that naturally satisfies a conservation equation. In
\cite{Wands:2000dp}, it was emphasized that the conservation, for
adiabatic perturbations and on large scales, of the linear
curvature perturbation on
 uniform-density hypersurfaces, usually denoted $\zeta$,
introduced in \cite{Bardeen:1983qw}, can be derived directly,
without resorting to Einstein's equations, from  the  conservation
of the energy-momentum tensor,
\beq
\label{conserv}
\nabla_a T^a_{\ b}=0.
\eeq
Here, we will use the same starting point in order to define a
non-linear generalization of $\zeta$, and we shall make use of the
covariant approach described in the previous section.

Let us  consider the projection along $u^a$ of the conservation
equation (\ref{conserv}). This yields
\beq
\label{continuity} \dot\rho + \Theta (\rho + P)=0,
\eeq
where we remind the reader that the dot represents a covariant
derivative projected along $u^a$, i.e. $\dot{} \equiv u^a\nabla_a$. If one
takes the projected gradient of the previous expression one gets
\beq
\label{first_rel}
D_a\left(\dot\rho\right)+\left(\rho+P\right)Z_a+\Theta
\left(X_a+Y_a\right)=0,
\eeq
where we have used the definitions (\ref{X_def}), (\ref{Y_def})
and (\ref{Z_def}).

We now concentrate on the first term of Eq.~(\ref{first_rel})
trying to invert the time derivative with the spatial gradient. In
order to do so, it will be  convenient to introduce the Lie
derivative along $u^a$ of tensors. The corresponding definition
for a covector, which will be useful below, is (see, e.g.,
\cite{wald} for the general definition)
\beq
\L \chi_a \equiv u^c \nabla_c \chi_a + \chi_{c} \nabla_a u^c. \label{Lie_def}
\eeq
Note that, for scalars, the dot derivation is equivalent  to the Lie
derivation, i.e. $\dot \rho = \L \rho$. Thus, we want to invert the Lie
derivation along $u^a$ with the spatial gradient in the first term
of Eq.~(\ref{first_rel}). The two do not commute and we have
\bea
D_a\left(\dot\rho\right)&=&h_a^{\
b}\nabla_b\left(u^c\nabla_c\rho\right)= \left(h_a^{\ b}\nabla_b
u^c\right) \nabla_c\rho +u^c h_a^{\ b}\nabla_b\nabla_c\rho \cr &
=&\L \left(D_a\rho\right)- \nabla_a u^c D_c \rho+\left(h_a^{\
b}\nabla_b u^c- u^b \nabla_b h_a^{\ c}\right) \nabla_c\rho.
\label{commutator}
\eea
This expression can be rewritten in a remarkably simple form as
\beq \label{rel_rho}D_a\left(\dot\rho\right) =
\L \left(D_a\rho\right)- \dot u_a \dot \rho.
\eeq

Although we have derived it for the energy density $\rho$, this
expression of course applies to any scalar quantity. In
particular, it can be applied to the integrated expansion
$\alpha$, thus allowing us to express $Z_a=D_a\Theta=3
D_a\dot{\alpha}$ in terms of $W_a=D_a\alpha$. This gives
\beq
\label{rel_alpha} \label{second_rel}
 Z_a=3\L W_a -3  \dot{u}_a \dot\alpha.
\eeq

Substituting (\ref{rel_rho}) and (\ref{rel_alpha}) in
(\ref{first_rel}), one gets
\bea
&& \L {X}_a+ 3\left(\rho+P\right) \L {W}_a+
\Theta\left(X_a+Y_a\right) =0, \label{eq17}
\eea
where terms proportional to $\dot{u}_a$ have disappeared because
of the continuity equation (\ref{continuity}). Using the relation
\beq
{{ \L {X}_a} \over \rho+P}= \L \left({X_a\over \rho+P}\right)-
\Theta (1+c_s^2){X_a\over \rho+P}, \label{deri}
\eeq
where we have introduced   $c_s^2 \equiv \dot P/ \dot\rho$, the
non-linear generalization of the adiabatic speed of sound,
Eq.~(\ref{eq17}) can be rewritten as an evolution equation for the
{\em covector}
\beq
\label{zeta} \zeta_a\equiv W_a +{X_a\over {3(
\rho+P)}}=D_a\alpha-\frac{\dot\alpha}{\dot\rho}D_a\rho,
\eeq
which, as we will see later, can be seen as a generalization of
the curvature perturbation on uniform density hypersurfaces of the
linear theory, usually denoted by $\zeta$ (hence our name
$\zeta_a$). Equation (\ref{eq17}) then yields
\beq
\L\zeta_a=-{\Theta\over{3(\rho+P)}}\left(Y_a -c_s^2 X_a\right) .
 \label{conserv1}
\eeq
On the right hand side,
 the quantity
 \beq
 \label{Gamma}
 \Gamma_a\equiv
Y_a -c_s^2 X_a=D_aP- {\dot P\over \dot\rho}D_a\rho
\eeq
can be interpreted as the projected gradient of the non-adiabatic
pressure, and represents the non-linear generalization of the
non-adiabatic pressure defined, e.g., in \cite{Wands:2000dp}. It
vanishes for purely adiabatic perturbations, i.e., when the
pressure $P$ is solely a function of the density $\rho$.

For practical purposes, it is  useful to note that, although both $\zeta_a$ and
$\Gamma_a$ are defined as linear combinations of {\it spatially projected gradients}, one can replace them
 by {\it ordinary gradients}, thus having
\beq
\label{partial}
\zeta_a=\partial_a\alpha-\frac{\dot\alpha}{\dot\rho}\partial_a\rho,
\qquad
\Gamma_a=\partial_a P- {\dot P\over \dot\rho}\partial_a\rho.
\eeq
This
is indeed possible for any linear combination of the form
$D_a\chi-(\dot\chi/\dot\eta)\D_a\eta$
since, for any scalar quantity $\chi$, we have
\beq
D_a \chi=\partial_a \chi+ u_a \dot {\chi}.
\eeq

Substituting the definition of $\Gamma_a$ into
Eq.~(\ref{conserv1}) we finally obtain
\beq
\L \zeta_a= -{\Theta\over{3(\rho+P)}} \Gamma_{a} .
\label{conserv2}
\eeq
This equation has a form very similar to the conservation equation
for $\zeta$ of the linear theory, which will be rederived in the
next section. However, in our case, the above equation  is
 {\it exact},  fully
{\it non-perturbative} and valid at {\it all scales}.

The time derivative that appears in the equation of the linear
theory has been replaced here by a Lie derivative along $u^a$,
which is consistent with the fact that we want to describe the
conservation properties of a {\em covector} along the fluid
worldline of a comoving observer. This makes Eq.~(\ref{conserv2})
remarkably simple and improves our original formalism given in
\cite{lv05a} where we did not introduce the Lie derivative.
Moreover, the use of the Lie derivative simplifies the
calculations in practice, because one does not need to compute the
covariant derivatives.  Indeed, the left hand side of
Eq.~(\ref{conserv2}) can be rewritten, starting from the
definition of Lie derivative (\ref{conserv2}), as
\beq
\L \zeta_a=  u^c
\partial_c \zeta_a +\zeta_c \partial_a u^c, \label{Lie_use}
\eeq
which will be  very simple to use in the following.

A remark is in order here. In our approach, the covector
$\zeta_a$, defined as a linear combination of spatial gradients of
$\alpha$ and of $\rho$, is the crucial quantity that enables us to
write the conservation equation for perturbations in a remarkably
simple form. It is interesting to note that the same linear
combination appears in the work of Rigopoulos and Shellard
\cite{Rigopoulos:2003ak}. Although their motivation is, like here,
to generalize the conservation equation for the linear $\zeta$ to
the non-linear case, their approach differs from ours in two
respects. First they do not use a covariant formalism but a
coordinate based approach with an ADM decomposition of the metric.
As a consequence, their correspondent equivalent of our $\alpha$
is not defined as the integrated expansion along fluid worldlines
but from the determinant of the spatial metric. In this, they
follow an older work \cite{Salopek:1990jq} which already
emphasized the advantage of this quantity to generalize the linear
$\zeta$ to non-linear order. Second -- and as a consequence of the
first point -- they restrict their analysis to super-Hubble scales
as in \cite{Salopek:1990jq}.

In a similar spirit, i.e., using the ADM decomposition and
invoking a long-wavelength approximation, the authors of
\cite{Lyth:2004gb} found, not exactly the linear combination
appearing in \cite{Rigopoulos:2003ak}, but its (spatially)
integrated version. In contrast with these previous works, as
already emphasized, our formulation is fully covariant and our
non-linear generalization applies to all scales.

\section{Linear theory} \label{Linear}

In this section, we compare our approach with the more familiar,
coordinate-based, linear theory (see e.g., \cite{liddle_lyth} or
\cite{cargese} for recent presentations of this topic). In order
to do so, we first introduce  conformal coordinates $x^\mu=\{\eta,
x^i\}$ to describe our almost-FLRW spacetime. A prime will denote
a partial derivative with respect to conformal time $\eta$, ${}' \equiv
\partial / \partial \eta$. 
The background spacetime, i.e., at zeroth order, is a
FLRW spacetime, endowed with the metric
\beq
ds^2={\bar g}_{\mu\nu}dx^\mu dx^\nu=a^2\left[-d\eta^2+
\gamma_{ij}dx^i dx^j\right],
\eeq
where $a=e^{{\bar\alpha}}$ is the background scale factor, and
filled with a homogeneous perfect fluid characterized by the
energy density ${\bar\rho}(\eta)$, the pressure ${\bar P}(\eta)$
and the four-velocity
\beq
{\bar u}^\mu=\{1/a,0,0,0\}.
\eeq
The perturbed spacetime is described by the perturbed metric
\beq
ds^2=\left({\bar g}_{\mu\nu}+\d g_{\mu\nu}\right)dx^\mu dx^\nu,
\eeq
with
\beq
\delta g_{00}=-2a^2A, \quad \delta g_{0i}=a^2B_i, \quad \delta
g_{ij}=a^2 H_{ij}.
\eeq
We decompose $H_{ij}$ in the form
\beq
\label{H}
H_{ij}=-2\psi \gamma_{ij}+2 \nabla_i\nabla_j E+2
\nabla_{(i}E^V_{j)}+2 { E}^T_{ij},
\eeq
with ${E}^T_{ij}$ transverse and traceless, i.e.,
$\nabla_i{E}^T{}^{ij}=0$ and $\gamma^{ij}{E}^T_{ij}=0$, and
$E^V_i$ transverse, i.e., $\nabla_iE^V{}^i=0$, where $\nabla_i$
denotes the covariant derivative with respect to the homogeneous
spatial metric $\gamma_{ij}$ (which is also used to lower or raise
the spatial indices).

The corresponding matter content is  a perfect fluid with
perturbed energy density and pressure,
\beq
\rho(\eta,x^i)={\bar\rho}(\eta)+\d \rho(\eta,x^i), \qquad
P(\eta,x^i)={\bar P}(\eta)+\d P(\eta,x^i)
\eeq
 and four-velocity
 \beq
 \label{u}
 u^\mu={\bar
u}^\mu+\d u^\mu, \quad \d u^\mu=\{-A/a, v^i/a\}, \quad v_k=
\nabla_k v+{ v}^V_k,
\eeq
where $v^V_i$ is transverse, $\nabla_i v^V{}^i=0$.

\def\Xf{X_{(1)}}
\def\Xs{X_{(2)}}
\def\af{\alpha_{(1)}}
\def\as{\alpha_{(2)}}
\def\rhof{\rho_{(1)}}
\def\pf{P_{(1)}}
\def\rhos{\rho_{(2)}}
\def\ps{P_{(2)}}
\def\ab{\bar{\alpha}}
\def\rhob{\bar{\rho}}
\def\pb{\bar{P}}

We now wish to write down explicitly the components of our vector
$\zeta_a$ in this generic coordinate system. We will start from
the expression for $\zeta_a$ given in (\ref{partial}) rather than
from its definition (\ref{zeta}). At zeroth order, i.e., in the
unperturbed FLRW spacetime, $\zeta_a$ automatically vanishes. At
{\it linear order},  the spatial components are simply
\beq
\zeta_i^{(1)}=\partial_i\zeta^{(1)}, \qquad \zeta^{(1)}\equiv
\af-{{\ab}'\over {\rhob}'}
\rhof,
\label{zeta1}
\eeq
where we recall that a prime denotes a partial derivative with
respect to conformal time.

To compute the component $\zeta_0$, it is useful to note that, for
any function $\chi$, one can write
\beq
D_0\chi=u_0 u^i\partial_i\chi-u^iu_i\partial_0 \chi,
\label{relation_chi}
\eeq
where we have used the normalization of $u^a$. Since $u^i$ is
first order, this implies that $\zeta_0^{(1)}=0$ (although this
will not be the case at second order).

In order to make the link with the usual quantities of the linear
theory, one needs to reexpress $\af$ in terms of the metric and
matter perturbations. The detailed calculations, up to second
order in the perturbations, are given in the appendix. Retaining
only the first order terms, we obtain
\begin{eqnarray}
\af&=&{1\over 6}\gamma^{ij}H_{ij}+\frac{1}{3} \int {d \eta}\,
\nabla_k v^k \cr &=& -\psi+{1\over 3}\nabla^2 E+ \frac{1}{3}\int
{d \eta}\, \nabla^2 v, \label{alpha}
\end{eqnarray}
where the decompositions (\ref{H}) and (\ref{u}) have been used to
obtain the second line.

The  components of the non-adiabatic term $\Gamma_a=\partial_a
P-(\dot P/\dot\rho)\partial_a\rho$ can be deduced directly from
the components of $\zeta_a$ by substituting $P$ to $\alpha$.
Therefore, one finds
\beq
\Gamma^{(1)}_i= \partial_i\Gamma^{(1)}, \qquad \Gamma^{(1)}\equiv
\pf-{{\pb}'\over {\rhob}'} \rhof, \label{Gamma1}
\eeq
while the time component vanishes.

Let us now specialize our equation (\ref{conserv2}) to first order
in perturbation theory. One can first notice that, at first
order, ${\cal L}_u \zeta^{(1)}_i = \zeta^{(1)}_i{}'/a$. We then
get
\beq
 \zeta_i^{(1)}{}'=-{\HH \over \rho+P}\Gamma^{(1)}_i, \label{lin1}
\eeq
where $\HH$ denotes the conformal Hubble rate, $\HH =
a'/a={\ab}'$. We then note that, from Eq.~(\ref{alpha}),
\beq
\af{}'=-\psi' +{1\over 3}\nabla^2 (E'+ v).
\eeq
Consequently, making use of Eqs.~(\ref{zeta1}) and (\ref{Gamma1})
and getting rid of the $\partial_i$ common to both sides of
(\ref{lin1}), one finds
\beq
\left(-\psi -\HH{\rhof\over\rhob'}\right)' +{1\over 3}\nabla^2(
E'+ v) =-{\HH\over \rho+P} \left(\pf-{{\pb}'\over {\rhob}'} \rhof
\right).
\eeq
One recognizes in the parenthesis of the left hand side the familiar quantity
\beq
\zeta\equiv-\psi -\HH{\rhof\over\rhob'}.
\eeq
Our linearized equation for $\zeta_a$ is thus shown to be
completely equivalent to the analogous equation obtained directly
in the linear theory
\beq
\zeta'=-{\HH\over \rho+P}\delta P_{\rm nad}-{1\over
3}\nabla^2\left(E'+ v\right), \label{lin2}
\eeq
with
\beq
\delta P_{\rm nad}\equiv \pf-{{\pb}'\over {\rhob}'} \rhof.
\eeq

As one can see, at the linear level, our quantity $\zeta_i$ is
conserved at {\it all scales} because our definition of $\zeta_i$
automatically includes the Laplacian terms that appear on the
right hand side of (\ref{lin2}). In other words, whereas our
$\zeta_a$ coincides with the usual $\zeta$ on long wavelengths
when the spatial gradients can be neglected, the two quantities
will differ on small scales, our quantity being more adapted to
the underlying conservation law.

\section{Second order perturbations}
In the following, we decompose any function $X$ in the form
\beq
X(\eta,x^i)={\bar X}(\eta)+ \Xf(\eta,x^i)+\frac{1}{2}
\Xs(\eta,x^i)+\dots \ ,
\eeq
where $\Xf$ and $\Xs$ represent, respectively,
the first and second order perturbations.

To compute the component $\zeta_0$ at second order we use again
Eq.~(\ref{relation_chi}). Since $u^i$ is first order, this implies
$\zeta_0^{(2)}=u^i\zeta_i^{(1)}$. Expanding now
$\zeta_i=\partial_i\alpha-(\dot\alpha/\dot\rho)\partial_i\rho$ up
to second order, one finds
\beq
\zeta_i^{(2)}=\partial_i\left( \as - {{\ab}'\over {\rhob}'}\rhos\right)
-{2\over \rhob'}\left(\af ' -{{\ab}'\over {\rhob}'} \rhof'\right)
\partial_i\rhof.
\eeq
In contrast with the first order expression, the coefficient in front
of the gradient in the second term on the right hand side cannot be directly
absorbed in the gradient because it depends on space. One can however
do the following manipulations:
\bea
\zeta_i^{(2)}&=&\partial_i\left( \as -{
{\ab}'\over  {\rhob}'} \rhos\right)
-{2\over {\rhob}'}\left(\af' -{
{\ab}'\over  {\rhob}'}\rhof' \right)
\partial_i\rhof
\nonumber \\
&=&
\partial_i \left(\as - \frac{\ab'}{\rhob'}\rhos
- 2 \af' \frac{\rhof}{\rhob'}     + 2
\ab'\frac{\rhof}{\rhob'}
\frac{\rhof'}{\rhob'} \right)
\nonumber
\\ &&+ 2 \partial_i \left( \af' -
\frac{\ab'}{\rhob'}
\rhof'
\right) \frac{\rhof}{\rhob'}
\nonumber \\
&=&
\partial_i \left[\as - \frac{\ab'}{\rhob'}\rhos
 - 2 \af' \frac{\rhof}{\rhob'}     + 2
\ab'\frac{\rhof}{\rhob'}
\frac{\rhof'}{\rhob'} \right. \nonumber
\\ && \left.+ \left(\ab'' - \ab' \frac{\rhob''}{\rhob'}
 \right) \frac{\rhof{}^2}{{\rhob'}{}^2}
\right]
 +  2 \frac{\rhof}{\rhob'} \partial_i  {
   \zeta^{(1)}}',
\eea
which can be written   in the form
\beq
\label{zeta2} \zeta_i^{(2)}=\partial_i\zeta^{(2)}+ {2\over
\rhob'}\rhof\partial_i{\zeta^{(1)}}',
\eeq
with
\beq
\zeta^{(2)} \equiv \as - {{\ab}'\over {\rhob}'}\rhos
 - \frac{2}{{\rhob}'} {\af}' \rhof+2 \frac{{\ab}'}{{{\rhob}'}{}^2} {\rhof}
 {\rhof}'+\frac{1}{{\rhob}'}
 {\left({{\ab}'\over {\rhob}'}\right)}'
 {\rhof}^2.
  \label{zeta_scal}
\eeq

If one substitutes  in the above equation the expression for
$\alpha$ up to second order derived in the appendix, keeping  only
the scalar terms without gradients, i.e., which are not negligible
on large scales,
\beq
\alpha \simeq \ln a - \psi - \psi^2 ,
\eeq
one finds
\beq
\zeta^{(2)} \simeq -\psi_{(2)} - 2\psi_{(1)}{}^2 -{{\HH}\over
{\rhob}'}\rhos
 + \frac{2}{{\rhob}'} {\psi_{(1)}}' \rhof+2 \frac{{\HH}}{{{\rhob}'}{}^2} {\rhof}
 {\rhof}'+\frac{1}{{\rhob}'}
 {\left({{\HH}\over {\rhob}'}\right)}'
 {\rhof}^2.
\eeq

The right hand side can be  easily related to the conserved second order
quantity defined by Malik and Wands in \cite{Malik:2003mv}, and thus,
\beq
\zeta^{(2)} \simeq \zeta^{(2)}_{\rm MW} - \zeta^{(1)2}_{\rm MW}.
\eeq
(See also the discussion in \cite{Lyth:2005du}.)

Now that we have identified the second order perturbation variable
$\zeta^{(2)}$, we can expand Eq.~(\ref{conserv2}) to second order.
On making use of Eq.~(\ref{Lie_use}) to reexpress the Lie
derivative along $u^a$ in terms of coordinate time derivative, and
retaining only second order terms, we have
\beq
\L \zeta^{(2)}_i = \frac{1}{a} \left[ \zeta^{(2)}_i{}' - 2 A
\zeta^{(1)}_i{}' +2 \left( v^j \partial_j \zeta^{(1)}_i +
\zeta^{(1)}_j
\partial_i v^j \right) \right].
\eeq
Finally, on making use of Eqs.~(\ref{zeta1}), (\ref{zeta2}), and
(\ref{zeta_scal}), and that $\Theta = 3 (1-A) \alpha'/a$ up to
first order, we can explicitly write the conservation equation
(\ref{conserv2}) at second order and {\em on all scales}:
\bea
  \zeta^{(2)} {}' &=& - \frac{\HH}{\bar \rho+ \bar P} \Gamma^{(2)} -
\frac{2}{\bar \rho+ \bar P} \Gamma^{(1)} \zeta^{(1)} {}'  - 2
v^j
\partial_j \zeta^{(1)} . \label{conserv_zeta2}
\eea
The definition of $\Gamma^{(2)}$ can be read from the expansion of
$\zeta^{(2)}$ by substituting $P$ to $\alpha$. For adiabatic
perturbations, we find that the second order scalar variable
$\zeta^{(2)}$ is conserved only on large scales, when the last
term on the right hand side of Eq.~(\ref{conserv_zeta2}) can be
neglected. In principle, one can  extend straightforwardly the
procedure presented here to higher orders in the perturbation
expansion.

\section{Other conserved quantities}
\subsection{Comoving curvature perturbation}
We have seen in the previous sections that our $\zeta_a$ can be
interpreted as the non-linear generalization of the curvature
perturbation on uniform energy density hypersurfaces of the linear
theory. Another useful quantity in the linear theory is the
so-called curvature perturbation on comoving hypersurfaces,
usually denoted by ${\cal R}$  and
one can wonder if a non-linear generalization can be constructed
within our formalism. As we have seen before, what plays the
r\^ole of the spatial curvature in our formalism is the integrated
expansion $\alpha$ and since our spatial gradients are defined
with respect to the comoving observers, a candidate that
generalizes ${\cal R}$ is simply
\beq
{\cal R}_a \equiv -D_a \alpha = -W_a.
\eeq

We can explicitly check the connection between this quantity
and the linear comoving curvature perturbation ${\cal R}$ by
noticing that, in the coordinate system  defined in the appendix,
one finds for the spatial components of ${\cal R}_a$, at first
order in the perturbations,
\beq
{\cal R}^{(1)}_i= -\partial_i \alpha^{(1)} - \dot{\bar \alpha} 
u_i = -\partial_i
\alpha^{(1)}- \bar \alpha' (v_i+B_i).
\eeq

Therefore, for the scalar part of ${\cal R}^{(1)}_i$, we recover
the usual definition of the comoving curvature perturbation,
\beq
{\cal R}_i^S{}^{(1)}=\partial_i {\cal R}^{(1)}, \quad \quad
{\cal R}^{(1)}=-\alpha^{(1)} - {\cal H}\left(v+B \right),
\eeq
 provided one can
replace $-\alpha^{(1)}$ by $\psi$.

It is relatively easy to derive an  evolution equation for ${\cal
R}_a$. As a first step, one can rewrite Eq.~(\ref{second_rel}) as
\beq
{\cal L}_u {\cal R}_a= -\frac{1}{3}\left( \Theta \dot{u}_a +Z_a
\right).
\label{step1}
\eeq
If one wishes to write the right hand side of this equation
only in terms of quantities characterizing the matter, there is
some extra work. This consists in replacing $Z_a$ with the use of
the field equations, i.e., Einstein's equations. Contracting the
identity
\beq
\nabla_c\nabla_du_a-\nabla_d\nabla_cu_a=R_{abcd}u^b,
\eeq
where $R_{abcd}$ is the Riemann curvature tensor, by
$g^{ac}h^{ed}$, one gets, upon using the decomposition
(\ref{decomposition}) and Einstein's field equations
[$R_{ab}=\kappa \left(T_{ab}-{1\over 2}T g_{ab}\right)$, 
with $\kappa =8 \pi G$], the so-called momentum constraint
equation (see, e.g., \cite{ellis}),
\beq
\frac{2}{3} Z_a - h_a^{\ b} \nabla_c (\sigma^c_{\ b} + \omega^c_{\
b}) + (\sigma^{\ b}_{a} + \omega^{\ b}_{a}) \dot u_b= - \kappa h_a^{\ b} T_{bc}u^c = 0,
\label{Za}
\eeq
where the right-hand side  vanishes for the energy-momentum
tensor (\ref{emt}).

Moreover, one can write $\dot u_a$ as
\beq
\dot u_a = - \frac{Y_a}{\rho+P}.
\label{euler}
\eeq
This is simply the Euler equation, which follows from
the projection, orthogonally to $u^a$,
 of the energy-momentum tensor conservation (\ref{conserv}).
The covector $Y_a$ can  be reexpressed as
\beq
\label{Y}
Y_a = \Gamma_a+ c_s^2 X_a.
\eeq
It is also worth noticing the identity that relates our two covectors
$\zeta_a$ and ${\cal R}_a$:
\beq
\zeta_a +{\cal R}_a={X_a\over 3(\rho+P)}.
\eeq

Finally, substituting the relations (\ref{Za}-\ref{Y})
 into Eq.~(\ref{step1}), we  obtain
\beq
{\cal L}_u {\cal R}_a = \left[\frac{\Theta \delta^{\ b}_a}{3(\rho+
P)} - \frac{\sigma^{\ b}_{a}+\omega^{\ b}_{a}}{2(\rho+P)} \right]
\left( \Gamma_b + c^2_s X_b \right)
 -\frac{1}{2} h_a^{\ b}
\nabla_c(\sigma^{c}_{\ b} + \omega^{ c}_{\ b}).   \label{Revol}
\eeq
This gives a fully non-linear evolution equation for ${\cal R}_a$.
As one can see, it is much more complicated than the evolution equation
for $\zeta_a$.

 At linear order, considering only scalar perturbations, 
one finds, from the definition  of $X_a$, 
\beq
X^{(1)}= \rho_{(1)} -3 \HH (\bar \rho+ \bar P)(v+B),
\eeq
where $X_i^S{}^{(1)}=\partial_iX^{(1)}$. The quantity $X^{(1)}$
corresponds to the so-called comoving density perturbation
\cite{Bardeen:1980kt}, known to vanish on large scales because of
the relativistic Poisson equation. Therefore, since $\zeta^{(1)} +
{\cal R}^{(1)}= X^{(1)}/[3 (\bar \rho +\bar P)]$, $-{\cal
R}^{(1)}$ coincides with $\zeta^{(1)}$ on large scales, and is
thus conserved on large scales. Note that this also holds at
second order, as shown in \cite{Vernizzi:2004nc}.

\subsection{The conserved variable $C_a$}
In the previous subsection, we  introduced
the variable ${\cal R}_a$ and showed that, in the linear limit and on large
scales, it reduces to the familiar comoving curvature perturbation.
This is however not the only non-linear quantity that satisfies this property.
As shown in \cite{Ellis:1989ju},
this is also the case for the projected gradient of the spatial
scalar curvature,
\beq
C_a= S^3 D_a \R. \label{Ca}
\eeq
To be more precise, the quantity $\R$ is defined as
\beq
\R = 2 \left(-\frac{1}{3} \Theta^2 + \sigma^2-\omega^2 + \kappa \rho
 \right),
\eeq
 with
$\sigma^2\equiv \frac{1}{2} \sigma_{ab} \sigma^{ab}$ and
$\omega^2\equiv \frac{1}{2} \omega_{ab} \omega^{ab}$. When
$\omega_{ab}=0$,  $\R$
 corresponds to the Ricci scalar curvature ${}^{(3)}\!R$ of
the three-dimensional hypersurfaces orthogonal to $u^a$. However,
in the general case when $\omega_{ab}\neq 0$, such hypersurfaces
cannot be defined, according to Froebenius' theorem (see, e.g.,
\cite{wald}).

As shown in \cite{Ellis:1989ju}, $C_a$ is conserved on large
scales in a linearly perturbed {\it flat} FLRW universe, for
adiabatic perturbations. More generally, it is convenient to
define the quantity \cite{EBH}
\beq
\tilde C_a = C_a - \frac{4 a K}{ \rho+P} X_a, \label{tildeCa}
\eeq
where $K=0,\pm1$ characterizes the spatial curvature of the
unperturbed FLRW spacetime. The variable $\tilde{C}_a$ reduces to
$C_a$ for a spatially flat background  and turns out to be
conserved on large scales for adiabatic linear perturbations for
all values of $K$.

To check the relation between $C_a$ and our quantity ${\cal R}_a$
at linear order, we now work with the perturbed metric  of
Sec.~\ref{Linear}. The first order perturbation of $\R$ is given
by \cite{Bruni:1992dg}
\beq
\R^{(1)} = \frac{4}{a^2} \left[ (\nabla^2 + 3 K) \psi - \HH
\nabla_k (v^k +B^k) \right].
\eeq
It follows from the definition (\ref{Ca}) that $C_0^{(1)}=0$ and, 
using $\overline {\cal K} = 6K/a^2$,
\bea
C_i^{(1)}&=&a^3 \left[(v_i+B_i) \partial_0 \overline{\R} +
\nabla_i \d \R^{(1)} \right] \nonumber \\ &=& 4 a \left\{ \nabla_i
\left[(\nabla^2 + 3K) \left(\psi - \HH (v+B)\right) \right] -3\HH
K (v^V_i+B^V_i) \right\}.
\eea
On making use of Eq.~(\ref{tildeCa}) one finds $\tilde
C_0^{(1)}=0$ and
\bea
\tilde C_i^{(1)} &=& 4 a \nabla_i \left\{\nabla^2 \left[ \psi -
\HH(v+B) \right]+3K\left[\psi - \frac{\rhof}{3(\rho+P)} \right]
\right\} \cr &=& 4 a \partial_i \left(\nabla^2 {\cal R}-3K \zeta
\right),
\eea
where $\zeta$ and ${\cal R}$ are the standard first order
perturbation variables.

\subsection{Conserved  number densities}

 We can define other types of conserved quantities starting
from any scalar quantity which obeys a continuity equation. In
analogy with the analysis of Lyth and Wands
 in the context of the linear theory \cite{Lyth:2003im}, one can, for
 example,
consider the particle number density in the physical contexts where the particle number
is conserved. The corresponding non-perturbative continuity equation is simply
\beq
\nabla_a\left(n \, u^a\right)=0,
\eeq
which yields
\beq
\dot n+\Theta n=0.
\label{conserv_n}
\eeq

Via a derivation similar to that of Sec.~3 for the energy density,
it is not difficult to show that the spatial projection of
(\ref{conserv_n}) can be rewritten as
\def\zetan{{\zeta^{(n)}}}
\beq
\L \zeta^{(n)}_a=0,
\eeq
where we have defined
\beq
\zeta_a^{(n)}\equiv W_a+{D_a n\over 3n}= D_a\alpha -
{\dot\alpha\over \dot n}D_a n.
\eeq
We thus see that, for any quantity that satisfies a standard local
conservation equation, one can construct a covector that embodies
a non-linear perturbation for this quantity, defined as a
combination of the spatial gradient of the quantity and that of
$\alpha$. This covector then satisfies an exact fully non-linear
conservation equation.

\section{Non-perturbative approach and gauge invariance}

The approach proposed in this work is completely non-perturbative:
it does not rely on a perturbative expansion of the variables
describing cosmological perturbations. This is why the equations
derived here encode the non-linear evolution of these variables.
In this section, we would like to comment about the notion of
gauge-invariance in the context of this approach.

Gauge-invariance has become ubiquitous in the studies of
linear cosmological perturbations. Indeed, using gauge-invariant
quantities helps avoiding ambiguities by getting rid of unphysical
degrees of freedom and makes  easier the comparison
between calculations done in different gauges. However, completely
fixing a gauge, either physically (i.e., with reference to some
specific matter) or geometrically, is equivalently acceptable.

In our case we do not really need to care about gauge-invariance
because we  use  tensor quantities that are physically well
specified independently of any coordinate system. Our approach,
however, contains one possible source of ambiguity: our quantity
$\alpha$ is defined up to a constant of integration along each
fluid worldline. Assuming that $\alpha$ is continuous implies that
there is some arbitrariness in the choice of the $\alpha=0$
hypersurface. A similar arbitrariness can be found in the separate
universe approach \cite{Sasaki:1995aw}.

This being said, it is nevertheless instructive to examine our
approach in the light of the treatment of gauge-invariance beyond
linear perturbations which has been discussed thoroughly in
\cite{Bruni:1996rg}. In the perturbative approach, one considers a
collection of space-times ${\cal M}_\lambda$ that interpolates
continuously, as the parameter
 $\lambda$ goes from $0$ to $1$, between the background spacetime ${\cal M}_0$ and the
 ``real'' universe ${\cal M}_1$. A choice of gauge corresponds to a choice of a continuous
 family of diffeomorphisms $\varphi_\lambda:{\cal M}_0\rightarrow {\cal M}_\lambda$. The perturbation
 of a tensor field $T$, which is associated to a continuous family of $T_\lambda$ defined on each
 ${\cal M}_\lambda$, is defined as
 \beq
 \Delta T_\lambda=\varphi^*_\lambda T- T_0,
 \eeq
 where $\varphi^*_\lambda T$ is the pull-back of $T_\lambda$. The perturbation $\Delta T_\lambda$ is thus
 a tensor field defined on ${\cal M}_0$. One can then do a Taylor expansion of the perturbation
 with respect to the parameter $\lambda$,
 \beq
 \Delta T_\lambda=\sum_{k=1}^\infty {\lambda^k\over k!}{\cal L}_\xi^k T,
 \eeq
 where $\xi$ is the tangent vector associated with the flow $\varphi_\lambda$, and one can thus define,
 order by order, the notion of gauge-invariance \cite{Bruni:1996rg}.

 In a non-perturbative approach, such as ours, one can define an alternative notion of gauge-invariance.
 Indeed one needs to consider only the background spacetime ${\cal M}_0$ and the
 real spacetime ${\cal M}$. A correspondence between ${\cal M}_0$ and ${\cal M}$ can be specified via
 a diffeomorphisms $\varphi:{\cal M}_0\rightarrow {\cal M}$. The gauge ambiguity results from
 the arbitrariness in the choice of $\varphi$. However, now, instead of defining the perturbation of
 a tensor $T$ on the background spacetime ${\cal M}_0$, one can do it on the perturbed spacetime:
 \beq
 \Delta T=T-\varphi_* T_0,
 \eeq
 where $\varphi_*$ is the push-forward of $T$. With this alternative
 definition for the perturbation, it is immediate  to see that $\Delta T$ is gauge-invariant,
 i.e.,
 independent of the choice of $\varphi$, if $T_0$ vanishes. In this sense, which does not coincide
 with the perturbative gauge-invariance defined in \cite{Bruni:1996rg} but which agrees implicitly with the
 discussion of  \cite{Ellis:1989jt}, the quantities defined in our approach are gauge-invariant.

\section{Conclusions}
In this work, we have discussed in detail  a new approach to the
theory of non-linear cosmological perturbations, which was
presented briefly in \cite{lv05a}.
 Our starting point is the covariant formalism for
cosmological perturbations and in particular its use of a
spatially projected gradient, defined orthogonally to the fluid
four-velocity. Our new formalism is then based on the following
additional ingredients:
\begin{itemize}
\item We have introduced $\alpha$, the local number of e-folds,
or integrated expansion,
along each worldline, and its spatially projected gradient.
\item We have introduced $\zeta_a$, a linear combination  of the spatial gradients of $\alpha$ and
of the energy density $\rho$. For {\it linear} perturbations and
on large scales, our definition reduces to the well-known
curvature perturbation on uniform energy density hypersurfaces
(more precisely, its gradient).
\item We have shown that $\zeta_a$ obeys a remarkably simple {\it conservation equation}, Eq.~(\ref{conserv2}), which is
{\it fully non-linear, exact and valid at all scales}.  Here,
improving the formulation of our previous work \cite{lv05a},
we describe the evolution of $\zeta_a$ by using the Lie
derivative with respect to the four-velocity $u^a$, instead of the
covariant derivative with respect to $u^a$. This explicitly shows
that $\zeta_a$ is {\em conserved, at all scales and fully
non-perturbatively}, for adiabatic perturbations. Indeed, the only
source of evolution of $\zeta_a$ is the non-linear generalization
of the non-adiabatic pressure, which describes entropy
perturbations and is defined as a linear combination of the
spatial gradients of $\rho$ and $P$.

\end{itemize}

We have shown explicitly the connection between our approach and
previous works, in particular on linear perturbations and on
second order perturbations. As an example, our equation (\ref{conserv2}) is so
simple that it is straightforward to write down the conservation
equation for the components of $\zeta_a$ at {\it second order} and
{\it at all scales}. One of the conclusions of this work is that
the local number of e-folds and the related combination $\zeta_a$
appear to be the most natural quantities to express the
conservation law for adiabatic cosmological perturbations, in its
fully non-linear and all-scale form.

As a final remark, let us stress that we did not use Einstein's
equations to derive the conservation equation for $\zeta_a$ but
only the conservation of the energy-momentum tensor. Therefore,
our equation (\ref{conserv2}) 
applies to {\it any} gravity theory, as long as the
energy-momentum tensor is conserved.

\vskip 1cm
{\bf Acknowledgments}
We  would like to thank Marco Bruni, Chris Clarkson,
Ruth Durrer, Kari Enqvist, Kazuya Koyama, David Lyth, Roy Maartens, Gerasimos
Rigopoulos, Paul Shellard, Bartjan W. van Tent,
and David Wands for useful discussions.
F.V. acknowledges
kind hospitality from the Institut d'Astrophysique de Paris.

\appendix

\section{Appendix: Perturbed quantities up to second order}
We consider a perturbed metric with
\beq
g_{00}=-a^2(1+2A), \quad g_{0i}=a^2B_i, \quad
g_{ij}=a^2\left(\gamma_{ij}+ H_{ij}\right),
\eeq
where $A$, $B_i$ and $H_{ij}$ are the perturbations. We 
decompose $H_{ij}$ in the form
\beq
H_{ij}=-2\psi \gamma_{ij}+2 \nabla_i\nabla_j E+2
\nabla_{(i}E^V_{j)}+2 {E}^T_{ij},
\eeq
with ${E}^T_{ij}$ transverse and traceless (TT), i.e.
$\nabla_i{E}^T{}^{ij}=0$ and $\gamma^{ij}{E}^T_{ij}=0$, and
$E^V_i$ is transverse, i.e. $\nabla_iE^V{}^i=0$. The components of
the inverse metric $g^{ab}$ are given, {\it up to second order},
by the following expressions:
\bea
&& g^{00}=\frac{1}{a^2} \left(-1+2A-4A^2 +B_i B^i \right), \quad
g^{0i}={1\over a^2}\left(B^i-2A B^i-H^{ik} B_k\right),\cr &&
g^{ij}={1\over a^2}\left(\gamma^{ij}-H^{ij}-B^i B^j+ H^i_{\
k}H^{kj}\right),
\eea
where the spatial indices are raised or lowered via the background
comoving spatial metric $\gamma_{ij}$. One has to imagine that each perturbation quantity can
be expanded up to second order, e.g., $A = A^{(1)}+\frac{1}{2}A^{(2)}$.

The time component $u^0$ of the four-velocity can be derived from
the normalization condition $g_{ab}u^au^b=-1$, and one finds {\it
at second order}
\beq
u^0=\frac{1}{a} \left(1-A+{3\over 2}A^2+{1\over 2}v_k v^k + B_k
v^k \right),
\eeq
where we have introduced $v^k\equiv a u^k$.

We now have all the ingredients to compute the expansion $\Theta=\nabla_a u^a$
{\it up to second order}. One finds
\bea
\Theta&=&\frac{1}{a} \left\{ (1- A)\left( 3 \HH+{1\over 2}
{H}^i_{\ i}{}' \right)-{1\over 2}H_{ij} {H}^{ij}{}' +{1\over
2}{\left[\left(v^k+B^k\right) \left(v_k+B_k\right)\right]}'
\right. \cr && \left.+3 \HH\left( {3\over 2} A^2 + {1\over
2}v_kv^k+ B_k v^k\right) +\nabla_k v^k +v^k \nabla_k \left(A+
\frac{1}{2} H^i_{\ i} \right) \right\} ,
\eea
where $H^i_{\ i}\equiv \gamma^{ij}H_{ij}= - 6 \psi + 2 \nabla^2
E$. This formula must be compared to the expression of $\Theta$ as
a function of the integrated expansion $\alpha$, $\Theta = 3 \dot
\alpha = u^a \nabla_ a \alpha$, which reads up to second order:
\beq
\Theta=\frac{3}{a} \left[ \left(1-A +{3\over 2}A^2+{1\over 2}v_k
v^k + B_k v^k\right) \alpha' +v^k\nabla_k \alpha \right].
\eeq
The order by order comparison finally yields
\bea
 {\alpha}'&=&\HH +\frac{1}{3} \left\{ {1\over 2}{ H}^i_{\ i}{}' -{1\over
2}H_{ij} {H}^{ij}{}' +{1\over 2}{\left[\left(v^k+B^k\right)
\left(v_k+B_k\right)\right]}' \right. \cr && \left. +(1+A)\nabla_k
v^k +v^k \nabla_k \left(A+\frac{1}{2}H^i_{\ i}\right) \right\} -
v^k \nabla_k\alpha^{(1)}.
\eea

\end{document}